\documentclass[a4paper,11pt]{article}
\usepackage{pos}
\usepackage{graphicx}
\usepackage{caption}
\usepackage{subcaption}

\title{Spectral clustering for jet reconstruction}

\author*[a]{Giorgio Cerro}
\author[b]{Srinandan Dasmahapatra}
\author[a,c,d]{Henry A. Day-Hall}
\author[a]{Billy Ford}
\author[a]{Shubhani Jain}
\author[a,e]{Stefano Moretti}
\author[c]{Claire H. Shepherd-Themistocleous}

\affiliation[a]{School of Physics and Astronomy, University of Southampton,\\
 	Southampton, SO17 1BJ, U.K.}

\affiliation[b]{School of Electronics and Computer Science, University of Southampton,\\
 	Southampton, SO17 1BJ, U.K.}
	
\affiliation[c]{Particle Physics Department, Rutherford Appleton Laboratory,\\
  	Chilton, Didcot, Oxon OX11 0QX, U.K.}
 
 \affiliation[d]{Faculty of Nuclear Sciences and Physical Engineering, Czech Technical University,\\
  	Prague, 160 00, Czech Republic}
	
\affiliation[e]{Department of Physics and Astronomy, Uppsala University, \\
	Uppsala, SE-751 20, Sweden}

\emailAdd{g.cerro@soton.ac.uk}
\emailAdd{sd@ecs.soton.ac.uk}
\emailAdd{henry.day-hall@cern.ch}
\emailAdd{b.ford@soton.ac.uk}
\emailAdd{s.jain@soton.ac.uk}
\emailAdd{stefano@phys.soton.ac.uk}
\emailAdd{claire.shepherd@stfc.ac.uk}

\abstract{
We present a new approach to jet definition alternative to clustering methods, such as the anti-$k_T$ scheme, that exploit kinematic data directly. Instead the new method uses kinematic information to represent the particles in a multidimensional space, as in spectral clustering. After confirming its Infra-Red (IR) safety, we compare its performance in analysing $gg \rightarrow H_{125~\rm GeV} \rightarrow H_{40~\rm GeV}H_{40~\rm GeV} \rightarrow b\bar{b}b\bar{b}$, $gg \rightarrow H_{500~\rm GeV} \rightarrow H_{125~\rm GeV}H_{125~\rm GeV} \rightarrow b\bar{b}b\bar{b}$ and $gg, q\bar{q} \rightarrow t\bar{t} \rightarrow b\bar{b}W^+W^- \rightarrow b\bar{b}jjl\nu l$ events from Monte Carlo (MC) samples, specifically, in reconstructing the relevant final states, to that of the anti-$k_T$ algorithm. Finally, we show that the results for spectral clustering are obtained without any change in the parameter settings of the algorithm, unlike the anti-$k_T$ case, which requires the cone size to be adjusted to the physics process under study.
}

\FullConference{%
  41st International Conference on High Energy physics - ICHEP2022\\
  6-13 July, 2022\\
  Bologna, Italy
}

\begin{document}
\maketitle

\section{Introduction}
In particle physics laboratories, when particles, at very high energies, collide,  an interesting and complex phenomena happens, the \textit{parton shower}. Immediately after the collision, partons (quarks and gluons), carrying lots of energy start interacting, producing more and more particles. Eventually, they lose some of their energy and combine themselves into hadrons, which are some of the particles that we are able to capture with detectors. Unfortunately, we do not have any access to what happens during the parton shower, but just to the final states particles, the ones that can be detected. One way to study what originated the shower is to cluster the particles trying to understand which jet they belong to, i.e. which is the common particle ancestor of each group.\\
Our main goal was to try improving the results obtained by the standard algorithms that physicists used for more than twenty years, such as anti-$k_T$, $k_T$ and Cambridge-Aachen. In order to tackle the problem we embraced Machine Learning (ML), which is becoming more and more popular nowadays. The model that we decided to develop is called \textit{Spectral Clustering}.

\section{Spectral clustering}
\begin{figure}
	\centering
	\includegraphics[width=15cm, height=4cm]{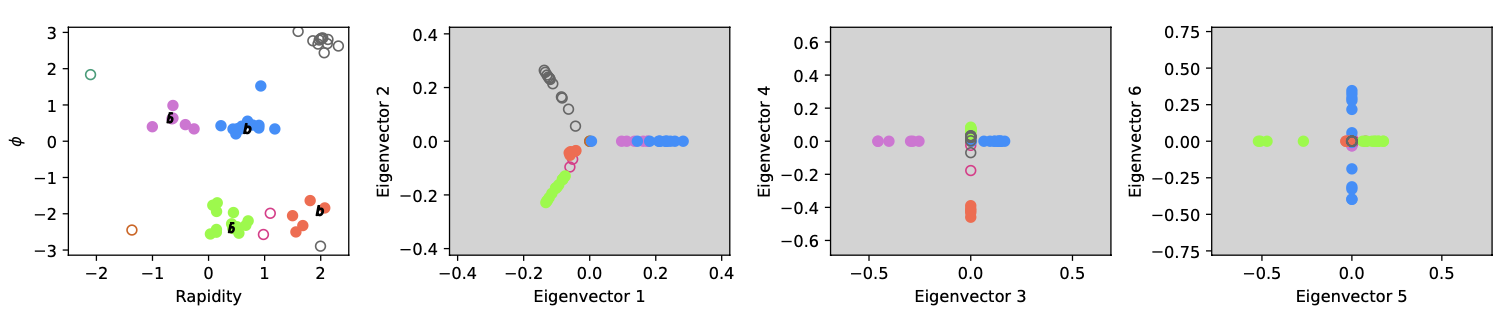}
	\caption{The left white plot shows the particles in the event as points on the unrolled detector barrel. The colour of each point indicates the jet it is assigned to, filled circles are b-jets. The three grey plots show the first 6 dimensions of the embedding space.}
	\label{emb_space}
\end{figure}

Spectral clustering is quite a common model for clustering, used in many different fields, such as movie suggestion and social networks,  but it is a novelty for particle physics. In this section we will present the key steps and the main concept of the algorithm adapted to our problem. \\
The very first step is to create a graph (an object with nodes, the particles, and edges, links between the nodes) and to encode the information of this graph into a symmetric matrix, the Laplacian, where all the rows and columns represent the particles, while the entries give the information of the connection between them. The next step is to perform the eigen decomposition, which returns as many eigen vectors and the number of particles. With these eigen vectors we create the \textit{embedding space}, a multidimensional space in which we project all the particles. It is interesting to see, fig.  \ref{emb_space}, that particles that belong to the same jet tend to align. This is a clear indication of the necessity of defining a new distance measure, the angular distance:
\begin{equation}
	d(t) _{i,j} = s(t)_{i,j} \arccos( \frac{m(t)_i \cdot m(t)_j}{||m(i)_i|| ||m(t)_j||})
\end{equation}
where $m(t)_i, m(t)_j$ are the particles' vectors in the new space, and $s(t)_{i,j}$ is a factor, whose values are in the range $[0,1]$ that ensures the Infra-Red and Collinear safety. 
This means that, ones the particles are embedded in the new space, all the pairwise angular distances are computed and the pair with the smallest distance is chosen to be merged: the two particles are removed from the set, replaced by the production of them, where the new particle carries a four-momentum equal to the sum of the two particles' four-momentum. All the steps (create the graph, embed the particles in the new space and merge the best pair) are repeated until a certain condition is met, i.e. the mean angular distance between all the particles reaches a certain threshold.

\section{Results}
To evaluate the behaviour of the spectral clustering method four datasets are used:
\begin{itemize}
 	\item \textit{Light Higgs}: a SM-like Higgs boson with a mass 125~\rm GeV decays into two light Higgs states with mass 40~\rm GeV, which in turn decay into $b\bar{b}$ quark pairs. The process is $gg \rightarrow H_{125~\rm GeV} \rightarrow H_{40~\rm GeV}H_{40~\rm GeV} \rightarrow b\bar{b}b\bar{b}$
	\item  \textit{Heavy Higgs}: a heavy Higgs boson with a mass 500~\rm GeV decays into two SM-like Higgs states with mass 125~\rm GeV, which in turn decay into $b\bar{b}$ quark pairs. The process is  $gg \rightarrow H_{500~\rm GeV} \rightarrow H_{125~\rm GeV}H_{125~\rm GeV} \rightarrow b\bar{b}b\bar{b}$ 
	\item \textit{Top}: a $t\bar{t}$ pair decays semileptonically, i.e. one $W^ {\pm}$ decayse into a pair of quark jets $jj$ and the other into a lepton-neutrino pair $l \nu_{l}$. The process is $gg, q\bar{q} \rightarrow t\bar{t} \rightarrow b\bar{b}W^+W^- \rightarrow b\bar{b}jjl\nu l$
\end{itemize}
The two measurements chosen to quantify the performance of the algorithms are the jet multiplicity and the mass peak.

\subsection{Jet multiplicity}
For all the events, the hard scattering produces four partons in the final states, which means that there are four jets that generate the shower. In clustering particles, ending up with four different clusters is a sign of good performance. We can appreciate the fact that spectral clustering obtained very well results, even outperforming the standard algorithms for the $Top$ case, fig. 

\begin{figure}[h]
	\centering
	\begin{subfigure}[b]{0.3\textwidth}
		\centering
		\includegraphics[width=5cm, height=4cm]{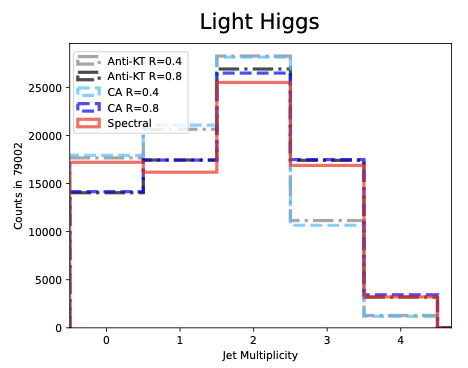}
	\end{subfigure} 
	\hfill
	\begin{subfigure}[b]{0.3\textwidth}
		\centering
		\includegraphics[width=5cm, height=4cm]{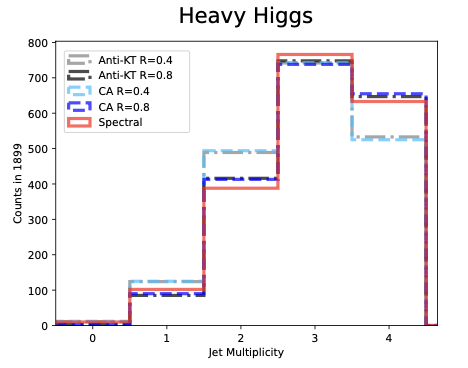}
	\end{subfigure}
	\hfill
	\begin{subfigure}[b]{0.3\textwidth}
		\centering
		\includegraphics[width=5cm, height=4cm]{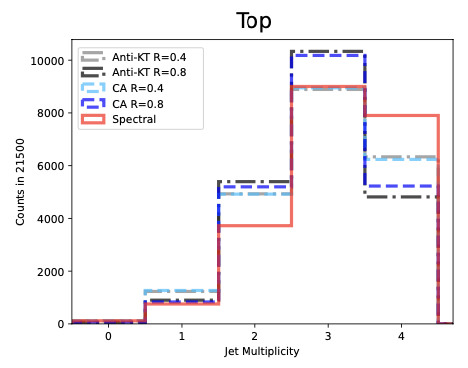}
	\end{subfigure}
		\caption{Jet multiplicities for the anti-$k_T$ and CA (for two cone radius choices) and spectral clustering algorithms on the \textit{Light Higgs}, \textit{Heavy Higgs} and \textit{Top} MonteCarlo samples. For all such datasets, spectral clustering performed at least as good as the other algorithms, obtaining the best results for the \textit{Top} case.}
		\label{multiplicities}
\end{figure}

\subsection{Mass peak}
\begin{figure}[h]
	\centering
	\includegraphics[width=15cm, height=5cm]{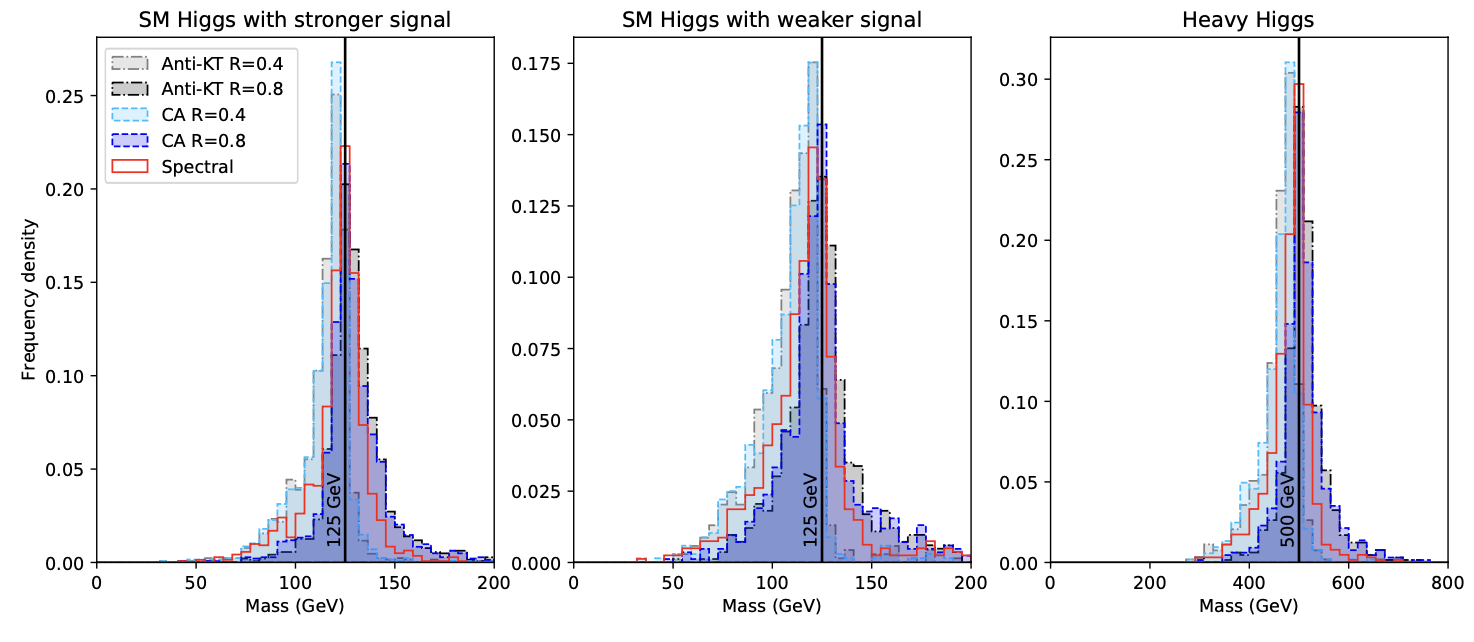}
	\caption{Three mass selections are plotted for the Light Higgs dataset. From left to right we show: the invariant mass of the 4b-jet system, of the 2b-jet system with heaviest with lightest invariant mass. Three jet clustering combinations are plotted as detailed in the legend. The spectral clustering algorithm is consistently the best performer in terms of the narrowest peaks being reconstructed and comparable to anti-$k_T$/CA.}
	\label{mass_peaks}
\end{figure}

The second measurement is the mass peak. Once the jets have been reconstructed, we look at the constituents of these jets and sum all their momenta to measure the overall invariant mass of the jets. A good algorithm centres the mean value to the the invariant mass of the particle that generated the shower, with a sharp distribution, i.e. a small invariance. From fig. \ref{mass_peaks}, we can see that, for the \textit{Heavy Higgs} case, spectral clustering performs as well as the standard algorithms, with sharp mass distribution centred at the right mean value. 

\section{Strengths and limits}
\begin{figure}[h]
	\centering
	\includegraphics[width=5cm, height=4.6cm]{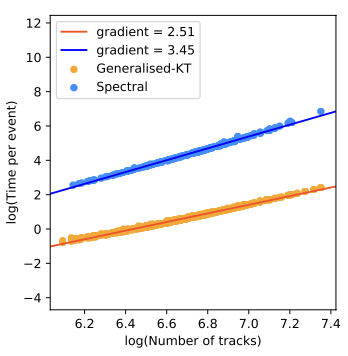}
	\caption{The run time of spectral clustering compared to a naïve implementation of generalised $k_T$ (without the performance refinements), on datasets of varying size. Simple fits are shown for each dataset in logarithmic scale. This shows that spectral clustering runs in just over $\textit{O}(n^3)$.}
	\label{speed}
\end{figure}

The great thing about spectral clustering, unlike most ML algorithms which work with black boxes, is that we have the total control over all the operation performed, which allows us to interact directly with the model, understanding in details all the steps.\\
The performance of ML models is also dictated by some tuneable parameters. Spectral clustering has just seven parameters and a grid search has been done in order to find the best set. What we found was that just a subset of these parameters actually had an impact on the performance of the model, i.e. some of them do not need to be tuned. Furthermore, we have been able to understand the exact role of each of these parameters in the model, which allow us to change them accordingly to what we require.  \\
Unfortunately, the biggest limit of spectral clustering is the speed. Performing the eigen decomposition of such big matrices is time and memory consuming, even more if it is required to do this operation at each time step. Looking at fig. \ref{speed} we can see that the complexity order of spectral clustering is about one order higher than the naïve implementation of $k_T$. 

\section{FastJet plugin}
We created a FastJet plugin. In this way, anyone who is curious about it and want to explore this new algorithm can download the code and use it immediately, as easy as it is to use the standard clustering algorithms, such as anti-$k_T$, $k_T$ and CA. \footnote{$https://github.com/HenryDayHall/fastjet \_spectraljet$}

\section{Coming next}
We are putting lots of effort on speeding it up, in order to make it at least competitive, in terms of speed, to the naïve versions of the standard algorithms. Moreover, we are investing in more detail all the properties of the embedding space, trying to understand whether there might be a more suitable way of merging, maybe even reconstructing the tree step by step.

\section{Acknowledgments}
We thank A. Chakraborty, J. Chaplais and E. Olaiya for insightful discussions. HAD-H thanks G. P. Salam for useful advice. HAD-H, SM and CHS-T are supported in part through the NExT Institute. SM is also supported by the STFC Consolidated Grant No. ST/L000296/1. We finally acknowledge the use of the IRIDIS High Performance Computing Facility, and associated support services, at the University of Southampton, in the completion of this work.

\nocite{*}
\bibliographystyle{plain} 
\bibliography{references}

\begin{thebibliography}{1}

\bibitem{Cacciari:2008gp}
Matteo Cacciari, Gavin~P. Salam, and Gregory Soyez.
\newblock {The anti-$k_t$ jet clustering algorithm}.
\newblock {\em JHEP}, 04:063, 2008.

\bibitem{Cacciari:2011ma}
Matteo Cacciari, Gavin~P. Salam, and Gregory Soyez.
\newblock {FastJet User Manual}.
\newblock {\em Eur. Phys. J. C}, 72:1896, 2012.

\bibitem{Cerro:2021abp}
Giorgio Cerro, Srinandan Dasmahapatra, Henry~A. Day-Hall, Billy Ford, Stefano
  Moretti, and Claire~H. Shepherd-Themistocleous.
\newblock {Spectral clustering for jet physics}.
\newblock {\em JHEP}, 02:165, 2022.

\bibitem{Chakraborty:2020vwj}
Amit Chakraborty, Srinandan Dasmahapatra, Henry Day-Hall, Billy Ford, Shubhani
  Jain, Stefano Moretti, Emmanuel Olaiya, and Claire Shepherd-Themistocleous.
\newblock {Revisiting jet clustering algorithms for new Higgs Boson searches in
  hadronic final states}.
\newblock {\em Eur. Phys. J. C}, 82(4):346, 2022.

\bibitem{Dokshitzer:1997in}
Yuri~L. Dokshitzer, G.~D. Leder, S.~Moretti, and B.~R. Webber.
\newblock {Better jet clustering algorithms}.
\newblock {\em JHEP}, 08:001, 1997.

\bibitem{vonLuxburg2007}
Ulrike von Luxburg.
\newblock A tutorial on spectral clustering.
\newblock {\em Statistics and Computing}, 17(4):395--416, Dec 2007.

\end{thebibliography}

\end{document}